\documentclass[12pt]{iopart}

\usepackage{iopams}
\usepackage{graphicx}

\begin{document}
\title[Strongly Interacting Atoms and Molecules in a 3D Optical Lattice]{Strongly Interacting Atoms and Molecules in a 3D Optical Lattice}
\author{Michael K\"ohl, Kenneth G\"unter, Thilo St\"oferle, Henning Moritz, and Tilman Esslinger}
\address{Institute of Quantum Electronics, ETH Z\"urich H\"onggerberg, CH-8093 Z\"urich, Switzerland}
\ead{koehl@phys.ethz.ch}

\begin{abstract}
We report on the realization of a strongly interacting quantum
degenerate gas of fermionic atoms in a three-dimensional optical
lattice. We prepare a band-insulating state for a two-component
Fermi gas with one atom per spin state per lattice site. Using a
Feshbach resonance, we induce strong interactions between the
atoms. When sweeping the magnetic field from the repulsive side
towards the attractive side of the Feshbach resonance we induce a
coupling between Bloch bands leading to a transfer of atoms from
the lowest band into higher bands. Sweeping the magnetic field
across the Feshbach resonance from the attractive towards the
repulsive side leads to two-particle bound states and ultimately
to the formation of molecules. From the fraction of formed
molecules we determine the temperature of the atoms in the
lattice.

\end{abstract}
\pacs{03.75.Ss, 05.30.Fk, 34.50.–s, 71.18.+y} \submitto{\JPB}
\maketitle

\section{Introduction}
Quantum degenerate atomic gases in optical lattices offer the
possibility to study quantum many-body physics with unprecedented
purity. Interacting atoms in optical lattices are ideally suited
to experimentally simulate the Hubbard model
\cite{Jaksch1998,Hofstetter2002}. For fermions this model is
elementary to describe the quantum physics of electrons in a
solid. It takes into account a single band of a static lattice
potential and assumes the interactions to be purely local
\cite{Hubbard1963}. The fundamental parameters include the tunnel
coupling between adjacent lattice sites, the atom-atom
interactions and the dimensionality of the system. The fermionic
Hubbard Hamiltonian reads
\begin{equation}
H=-t\sum_{<j,l>,\sigma} c^\dag_{j,\sigma} c_{l,\sigma} + U \sum_j
n_{j,\uparrow}n_{j,\downarrow}. \label{hubbardhamiltonian}
\end{equation}
where $t$ is the hopping matrix element and $c^\dag_{j,\sigma}$
and $c_{j,\sigma}$ are the creation and annihilation operators for
a particle in the spin state $\sigma$ at lattice site $j$,
respectively. $n_{j,\sigma}=c^\dag_{j\sigma}c_{j,\sigma}$ is the
number operator at lattice site $j$, and $U$ quantifies the
strength of the on-site interaction between atoms in different
spin states.

Despite its conceptual simplicity the Hubbard Hamiltonian has not
been solved except in the one-dimensional situation and in very
few special cases in higher dimensions. Therefore, experimentally
simulating the Hubbard model with ultracold atomic gases in
optical lattices is a fascinating possibility to explore the phase
diagram of this model. Both the tunneling matrix element $t$ and
the on-site interaction $U$ can be adjusted experimentally.
Whereas the tunnelling mainly depends on the depth of the periodic
potential, the on-site interaction is determined by the s-wave
scattering length $a$ for a collision between two distinguishable
fermions. The s-wave scattering can be modified by means of a
magnetically induced Feshbach resonance \cite{Inouye1998} which
allows to access any value of the scattering length.

As a first step along the path of such an experimental simulation
we study the Hubbard model in the low-tunneling regime, where $t$
is small and particles are assumed to stay on their initial
lattice sites for the duration of the experiment. In this
configuration approximative solutions of this model are feasible
\cite{Carr2005,Zhou2005,Diener2006,Katzgraber2005}. In the Hubbard
model the interactions between particles are parametrized only by
the parameter $U$ regardless of the physical details of the
interaction. In reality, however, the interatomic van-der-Waals
interactions are much more complex and the relevance of these
details for the Hubbard model must be investigated before unknown
many-body quantum phases can be addressed. Moreover, the
possibility of converting fermionic atoms into bosonic molecules
offers new insights beyond what has been studied in condensed
matter physics so far. The physics of two interacting particles in
a tight harmonic trap is governed by several length scales. These
are the characteristic length of the van-der-Waals interaction
potential between the atoms, the effective range, the scattering
length and the extension of the ground state in the oscillator
potential.

\section{Tuning Interactions Between Atoms}
\label{sec:theory_fermions_feshbach}

In an atomic gas, interactions are mediated by scattering
processes. Because of the Pauli principle, elastic s-wave
scattering in a degenerate fermionic gas takes place only in spin
mixtures or mixtures with other atomic species. Feshbach
resonances, a phenomenon originally discussed in the context of
nuclear physics \cite{Feshbach1958}, allow a tuning of the
scattering properties in a binary collision between two atoms to
arbitrary repulsive or attractive values \cite{Tiesinga1993}. This
enables the creation of strongly interacting atomic quantum gases.

\subsection{A Simple Model: Two Interacting Atoms in a Harmonic
Oscillator Potential}

The collision properties between atoms are modified if the atoms
are subject to strong confinement. If the particles are confined
to a length scale comparable to the scattering length between
them, the known picture of two-particle bound states has to be
reconsidered. Such a tight confinement can for example be realized
in a three-dimensional optical lattice, where atoms are localized
to the lattice sites and the interactions can be tuned by a
Feshbach resonance.

\begin{figure}[htbp]
    \begin{center}
        \includegraphics[width=.6\columnwidth,clip=true]{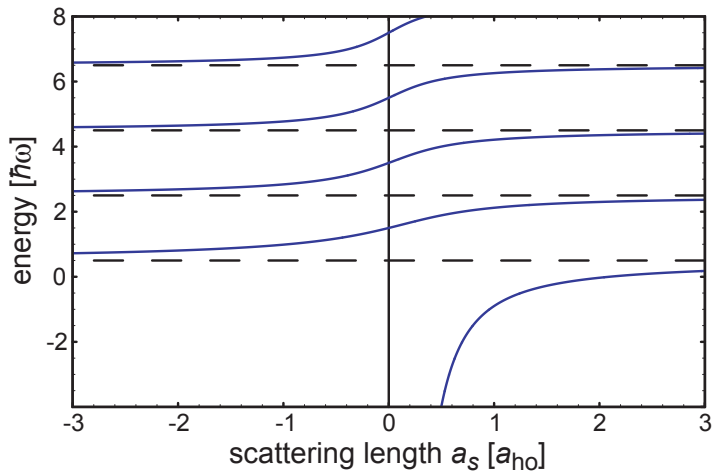}
        \caption{Energy spectrum of two interacting particles in a three-dimensional harmonic oscillator potential
        which is given by equation (\ref{eqHO}) in the center-of-mass frame \cite{Busch1998}.
        The asymptotic energies at $a_s \rightarrow \pm\infty$ are indicated by the dashed lines.}
        \label{fig1}
    \end{center}
\end{figure}

For simplicity we approximate a single potential well of the
lattice by a three-dimensional harmonic oscillator potential. This
fundamental quantum mechanical model system has been studied
theoretically and the eigenenergies have been calculated in
various approximations
\cite{Busch1998,Blume2002,Bolda2002,Dickerscheid2005}. The
interaction between two particles in a harmonic oscillator affects
only the relative motion between the particles. After separating
off the center-of-mass degree of freedom we find the Hamiltonian
for the relative motion
\begin{equation}
H_{rel}=-\frac{\hbar^2 \nabla^2}{2\mu}+\frac{1}{2}\mu \omega^2
r^2+V(r)
\end{equation}
with the reduced mass $\mu=m/2$, the atomic mass $m$, the trapping
frequency $\omega$ and the two-particle interaction potential
$V(r)$. We assume s-wave interactions with a scattering length
$a$, which can be modelled by a regularized pseudopotential
interaction
\begin{equation}
V(r)=\frac{4 \pi \hbar^2
a}{m}\delta^{(3)}(r)\frac{\partial}{\partial r}r.
\end{equation}
This problem can be solved analytically \cite{Busch1998} and the
eigenenergies are given by the implicit equation
\begin{equation}
\frac{a_{ho}}{a}=\sqrt{2}\frac{\Gamma(-E/2\hbar
\omega+3/4)}{\Gamma(-E/2\hbar \omega+1/4)}, \label{eqHO}
\end{equation}
where $a_{ho}=\sqrt{\hbar/m \omega}$ and $\Gamma(x)$ denotes the
Gamma function. In figure \ref{fig1} we show the eigenenergies as
a function of the scattering length. For $|a|/a_{ho}\ll 1$ the
deviation of the eigenenergy from the noninteracting system is
linear in $a$, as one would have expected naively. However, for
large scattering length the energy shift saturates at $\pm \hbar
\omega$ with respect to the noninteracting case.

In reality, the atoms are interacting via a van-der-Waals
potential which is not at all $\delta$-function like, but has a
tail which drops off as $1/r^6$ for large distances. Naturally the
question arises, whether the pseudopotential approximation is
valid for the case of two atoms. The validity of the
pseudopotential approximation depends on the various length scales
involved in the problem. Generally speaking, a pseudopotential
treatment is valid, as long as the characteristic range of the
van-der-Waals interaction $\beta_6=(2 \mu C_6/\hbar^2)^{1/4}$ is
small compared with $a_{ho}$ \cite{Blume2002,Bolda2002}. This
assures that the interatomic potentials are not modified by the
confinement and the microscopic (i.e. atomic) physics is not
altered. For our parameters with $\beta_6/a_{ho}<0.1$ this is well
fulfilled. In \cite{Blume2002,Bolda2002} it is argued that one has
to consider an energy-dependent pseudopotential, if the effective
range $r_\textrm{eff}$ of the interaction becomes comparable to
the scattering length $a$ or the ground state extension $a_{ho}$.
Then the scattering length takes the form
\begin{equation}
a(E)=\left(\frac{1}{a}-\frac{1}{2a_{ho}}\frac{E}{\hbar
\omega}\frac{r_\textrm{eff}}{a_{ho}} \right)^{-1}.
\end{equation}
The effective range can for the case of $^{40}$K be calculated to
be \cite{Gao1998}
\begin{equation}
r_\textrm{eff}\simeq\frac{\beta_6}{3}\left(-\frac{4
\beta_6}{a}+\frac{8 \pi \beta_6^2}{a^2
\Gamma(\frac{1}{4})^2}+\frac{\Gamma(\frac{1}{4})^2}{\pi}\right)=
98 a_0,
\end{equation}
which is the same order of magnitude as $\beta_6$ and $a_{bg}$.

\section{Interacting Fermions in a Lattice}

For strong interactions between the fermions a multi-band Hubbard
model must be considered which takes into account a coupling
between Bloch bands \cite{Diener2006} and the conversion of pairs
of fermionic atoms into bosonic molecules. The regime of
Fermi-Bose conversion is not accessible in standard condensed
matter systems and only recently the first steps to understand
this mixed world of fermions and bosons have been undertaken
theoretically \cite{Dickerscheid2005,Carr2005,Zhou2005}. However,
multi-band Hubbard models are extremely difficult to solve
\cite{Troyer2005} and often solutions can only be derived for the
low-tunnelling limit. In this limit the lattice can be considered
as an array of microscopic harmonic traps occupied with two
interacting atoms in different spin states. In the limit of zero
tunneling the lowest Bloch band maps onto the lowest vibrational
harmonic oscillator state and higher bands map to excited
oscillator states. This allows us to apply the results from the
previous paragraph to the optical lattice. Feshbach resonances
\cite{Feshbach1958} allow for a controlled manipulation of the
atomic scattering properties by means of an external magnetic
field. Sweeping a magnetic field dynamically changes the two-body
potentials and the adiabatic creation and dissociation of
molecules can be observed, depending on the direction the magnetic
field sweep.

\subsection{Interaction induced coupling between Bloch bands}

\begin{figure}[ht]
  \begin{center}
  \includegraphics[width=.7\columnwidth,clip=true]{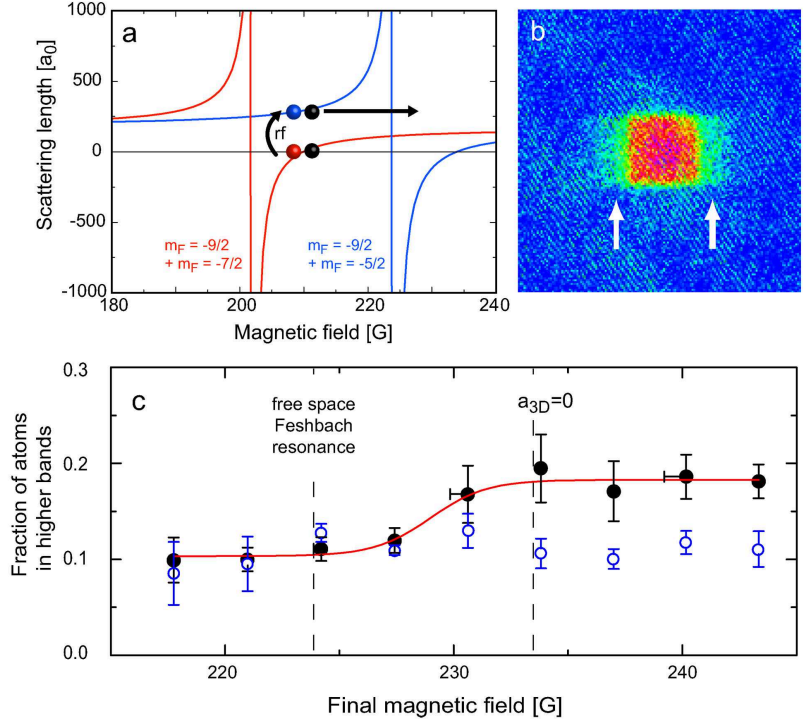}
  \end{center}
  \caption{Interaction-induced transition between Bloch bands. (a) Illustration of
  the sweep across the Feshbach resonance. (b)  Momentum distribution for
a final magnetic field of $B=233$\,G and a $12\,\mu$s/G sweep
rate. Arrows indicate the atoms in the higher bands. (c)
Transferring fermions to higher bands using a sweep across the
Feshbach resonance (filled symbols). The inverse magnetic field
sweep rate is $12\,\mu$s/G. The line shows a sigmoidal fit to the
data. The open symbols show a repetition of the experiment with
the atoms prepared in the spin states $|m_F=-9/2\rangle$ and
$|m_F=-7/2\rangle$ where the scattering length is not very
sensitive to the magnetic field. The magnetic field is calibrated
by rf spectroscopy between Zeeman levels. Due to the rapid ramp
the field lags behind its asymptotic value and the horizontal
error bars represent this deviation. Data taken from
\cite{Kohl2005b}.}
  \label{fig2}
\end{figure}

We have investigated the behaviour of the atoms when sweeping
across the Feshbach resonance from the low-field to the high-field
side (see figure \ref{fig2}a) \cite{Kohl2005b}. When using this
direction of the sweep there is no adiabatic conversion to
molecules, but a transfer of atoms into an excited vibrational
level of the harmonic oscillator is expected. We start from a
noninteracting gas deep in a band insulator regime with
$V_x=12\,E_r$ and $V_y=V_z=18\,E_r$ and corresponding trapping
frequencies of $\omega_x=2 \pi \times 50$\,kHz and
$\omega_y=\omega_z=2 \pi \times 62$\,kHz in the individual
potential minima. $E_r=h^2/(2 m \lambda^2)$ denotes the recoil
energy for a lattice laser with wave length $\lambda$. We prepare
a mixture of the two atomic states $|F=9/2, m_F=-9/2\rangle$ and
$|F=9/2, m_F=-5/2\rangle$, which have a Feshbach resonance at a
magnetic field of 224\,G \cite{Regal2003b}. In the following we
will refer to the atomic states only by the $m_F$ quantum number.
Starting from the initial magnetic field of 210\,G we ramp the
magnetic field with an inverse sweep rate of $12\,\mu$s/G to
different final values around the Feshbach resonance. After
turning off the optical lattice adiabatically and switching off
the magnetic field we measure the momentum distribution which
reflects the quasi-momentum distribution of the atoms in the
lattice \cite{Greiner2001b,Kohl2005b}. The adiabatic switch-off
process of the lattice adiabatically converts the quasi-momentum
of the atoms in the lattice into momentum which can be observed in
a time-of-flight image. To study the effect of the interactions we
determine the fraction of atoms transferred into higher bands,
which corresponds to excited states inside the individual lattice
wells. For final magnetic field values well above the Feshbach
resonance we observe a significant increase in the number of atoms
in higher bands along the weak axis of the lattice, demonstrating
an interaction-induced coupling between the lowest bands. The
fraction of atoms transferred could be limited by the number of
doubly occupied lattice sites and tunnelling within the higher
bands. The fraction of doubly occupied lattice sites was
determined in a different experiment to be approximately $40\%$
\cite{Stoferle2005b}. A recent theoretical work suggests that up
to $75\%$ of these atoms could be transferred into higher bands,
which would result in a total fraction of $30\%$ in higher bands
\cite{Diener2006}, which is somewhat more than observed in our
experiment.

\subsection{Producing bosonic molecules from fermionic atoms}

\begin{figure}[htbp]
\begin{center}
  \includegraphics[width=.6\columnwidth,clip=true]{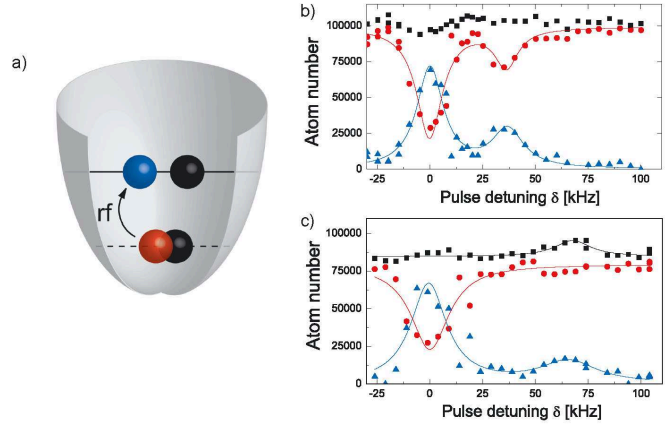}
  \end{center}
  \caption{Illustration of the rf spectroscopy between two bound states
  within the wells of the optical lattice. {\bf a)} The atoms in the initial states $|-7/2\rangle$ (red) and
  $|-9/2\rangle$ (black) are converted into a bound dimer by sweeping
  across a Feshbach resonance if lattice sites are doubly occupied.
  Application of an rf pulse on the transition $|-7/2\rangle \rightarrow
  |-5/2\rangle$. For singly occupied sites the rf resonance is at the bare atom transition given by the magnetic field. If a site is
  doubly occupied the rf transition dissociates the molecule. For this the binding energy $E_B$ has to be supplied
  and the rf resonance is shifted.
  {\bf b)} rf spectrum taken at $B=202.9$\,G, i.e. for $a<0$, and a lattice depth of 22\,$E_r$.
  The atom numbers are shown for the $|-9/2\rangle$ (squares),
  $|-7/2\rangle$ (circles), and $|-5/2\rangle$ (triangles) states. The lines are Lorentzian fits to the data. Data taken from \cite{Stoferle2005b}.}
  \label{fig3}
\end{figure}

We now study the reverse sweep direction where pairs of atoms are
transferred into deeply bound states \cite{Stoferle2005b}. This
technique of adiabatic conversion of atoms into molecules
\cite{Regal2003b} provides a useful way to generate ultracold
molecular samples. The lowest branch of the energy spectrum in
figure \ref{fig1} is accessed by starting from a noninteracting
Fermi gas but sweeping the magnetic field from the high-field
towards the low-field side of the Feshbach resonance. This sweep
adiabatically converts the atomic pairs into molecules. First we
create a band insulator  for each of the two fermionic spin states
$|-7/2\rangle$ and $|-9/2\rangle$ in the optical lattice.
Subsequently, the molecules are formed by ramping the magnetic
field from the zero crossing of the scattering length at
$B=210$\,G in 10 ms to its desired value close to the Feshbach
resonance located at $B_0=202.1$\,G \cite{Regal2004a}. We measure
the binding energy $E_B$ of the dimers by radio-frequency
spectroscopy \cite{Regal2003b,Chin2004,Moritz2005}. The idea of
the rf-spectroscopy is shown schematically in figure \ref{fig3}a:
An rf pulse dissociates the molecules and transfers an atom from
the state $|-7/2\rangle$ into the initially unpopulated state
$|-5/2\rangle$ which does not exhibit a Feshbach resonance with
the state $|-9/2\rangle$ at this magnetic field. Therefore the
fragments after dissociation are essentially noninteracting and
occupy the non-interacting ground state of the harmonic oscillator
potential. We vary the detuning $\delta=\nu_{RF}-\nu_{0}$ of the
rf pulse from the resonance frequency $\nu_0$ of the atomic
$|-7/2\rangle \rightarrow |-5/2\rangle$ transition. The power and
the duration of the pulse are optimized to constitute
approximately a $\pi$-pulse on the free atom transition. The
number of atoms in each spin state is detected using absorption
imaging after ballistic expansion.

Figure \ref{fig3}b shows rf spectra of atoms and molecules trapped
in a three-dimensional lattice with a potential depth of
$V_0=22\,E_r$ corresponding to $\omega=2 \pi \times 65$\,kHz. The
spectrum in figure \ref{fig3}b is taken at a magnetic field of
$B=202.9$\,G where $a/a_{ho}=-1.3$. For negative scattering
lengths, the molecules are only bound when they are strongly
confined whereas no bound state would exist in the homogeneous
case \cite{Wigner1933}. The spectrum exhibits two resonances: the
one at $\delta=0$ corresponds to the atomic transition from the
$|-7/2\rangle$ into the $|-5/2\rangle$ state. This transition
takes place at all lattice sites which initially were only singly
occupied and no molecule can be formed. The second resonance at
$\delta>0$ corresponds to molecule dissociation and is shifted
from the atomic resonance by the binding energy. Together with the
increase in the $|-5/2\rangle$ atom number we observe a loss of
atoms in the $|-7/2\rangle$ state, whereas the $|-9/2\rangle$
remains unaffected. This is expected since the rampdown of the
lattice before detection dissociates all molecules and the
$|-9/2\rangle$ atom number should be fully recovered.

In contrast to earlier work on rf dissociation of molecules where
the molecules were dissociated into a continuum and the fragments
were essentially free particles
\cite{Regal2003b,Chin2004,Moritz2005} in our configuration the
fragments occupy a discrete energy eigenstate of the confining
potential. In such a bound-bound transition no extra kinetic
energy is imparted onto the dissociated fragments and we determine
the binding energy from the separation of the atomic and the
molecular peak. Moreover, since there is at most one molecule
present per lattice site, collisional shifts
\cite{Harber2002,Gupta2003} are absent and we can estimate the
error in the binding energy from the fit error which is less than
5\,kHz.


\begin{figure}[htbp]
\begin{center}
  \includegraphics[width=.5\columnwidth,clip=true]{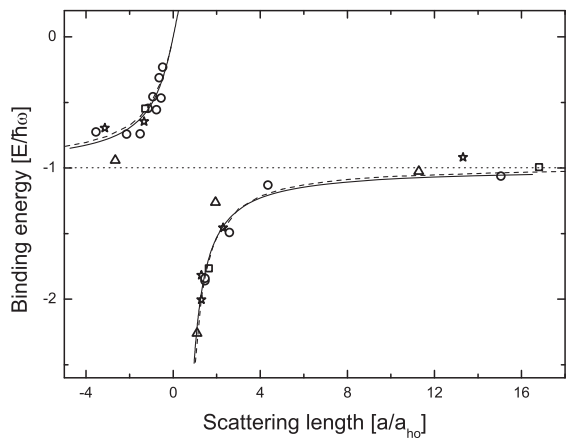}
\end{center}
  \caption{The measured binding energy of molecules in a three-dimensional optical lattice. The data are taken for
  several potential depths of the optical lattice of 6\,$E_r$ (triangles), 10\,$E_r$ (stars), 15\,$E_r$ (circles), and 22\,$E_r$
  (squares). The solid line corresponds to the theory of ref. \cite{Busch1998} with no free
  parameters, the dashed line uses an energy-dependent
  pseudopotential according to ref. \cite{Blume2002}.
  At the position of the Feshbach resonance ($a\rightarrow \pm \infty$, horizontal dashed line)
  the binding energy takes the value $E=-\hbar \omega$. Data taken from \cite{Stoferle2005b}.}
  \label{fig4}
\end{figure}

We have investigated the dependence of the binding energy of the
molecules on the scattering length (Fig.\;\ref{fig4}). The
scattering length is derived from the magnetic field using the
parametrization of the Feshbach resonance
$a(B)=a_{bg}(1-\frac{\Delta B}{B-B_0})$, with $a_{bg}=174\,a_0$
\cite{Regal2003a} and $\Delta B=7.8$\,G \cite{Regal2004b}. We
compare our data with the theory for the binding energy of two
particles trapped in a harmonic oscillator potential interacting
via an energy-independent pseudopotential \cite{Busch1998}.
Depending on the scattering length the binding energy $E$ of the
molecules varies according to equation (\ref{eqHO}). We have
determined the ground state extension $a_{ho}$ by minimizing the
energy of a Gaussian trial wave function inside a single well of
our lattice potential. We find the normalized binding energy
$E/\hbar \omega$ to be independent of the strength of the lattice
and all data points to agree well with the theoretical prediction
of equation (\ref{eqHO}) without adjustable parameters. However, a
model with an energy-dependent pseudopotential \cite{Blume2002}
gives very similar results (dashed line in figure \ref{fig4}).
Both models agree to within a few percent, which is small compared
to experimental uncertainties. Further improvements taking into
account more details of the atom-atom interaction in a two-channel
model have been suggested \cite{Bolda2002,Dickerscheid2005} and
could be tested with our data.

\section{Thermometry in the optical lattice}

In contrast to conventional condensed matter physics experiments
atoms in optical lattices are subject to an inhomogeneous
potential. This makes the concept of the ''filling'' of the
lattice more subtle than in the homogeneous case. One possible way
to quantify the filling of the inhomogeneous lattice is to study
the fraction of doubly occupied sites for a two-component Fermi
gas. The fraction of doubly occupied lattice sites is closely
related to the temperature of the atoms in the lattice and
therefore it represents a useful way of determining the
temperature of atoms in the lattice. This quantity is of
importance for many proposed experiments in which phase
transitions of fermions in lattices are studied, and special
cooling techniques to reach these temperatures have been devised
\cite{Hofstetter2002,Rabl2003,Blakie2005,Werner2005}.

\begin{figure}[htbp]
\begin{center}
  \includegraphics[width=.5\columnwidth,clip=true]{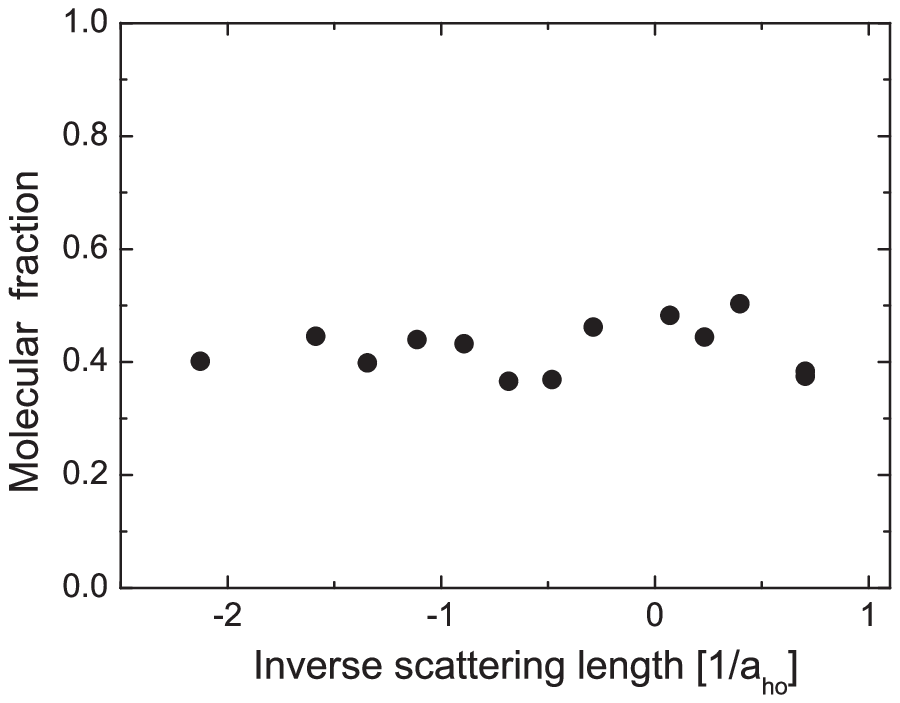}
\end{center}
  \caption{The measured molecular fraction as a function of the inverse scattering length at the end of the magnetic
  field sweep which is derived from the peak heights of the rf
  spectra and corrected by the overlap intergral. We find a constant molecular fraction of $(43\pm5)\%$.}
  \label{fig5}
\end{figure}

In the tight binding regime molecules can only be formed, when two
particles reside at the same lattice site. Therefore the
measurement of the molecule formation gives access to the fraction
of doubly occupied lattice sites. We assume that the magnetic
field sweep across the Feshbach resonance is fast enough such that
the atomic density does not change and we form molecules at all
those lattice sites where two atoms reside. Using the rf
dissociation technique we break up the molecules and determine the
fraction of molecules from the spectra. The dissociation
probability depends on the strength of the rf coupling, on the
overlap between the initial and the final wave function and on the
fraction of doubly occupied lattice sites. When we compare the
spin-flip probability on the atomic transition and on the
molecular transition, the rf coupling strength is equal, but the
spatial overlap is different. For the atomic transition, the
overlap is approximately unity since both, initial and final state
correspond to the noninteracting ground state of the harmonic
potential well. For the molecular peak we calculate the overlap
integral
\begin{equation}
 \left|\int d^3r\psi_{ho}({\bf r})^*\psi_{l=0}({\bf r}) \right|^2
\end{equation}
with $\psi_{ho}=(\pi a_{ho}^2)^{-3/4}e^{r^2/2 a_{ho}^2}$ being the
oscillator ground state wave function. $\psi_{l=0}({\bf
r})=\frac{1}{2}\pi^{-3/2}A e^{-r^2/2
a_{ho}^2}\Gamma(-\nu)U(-\nu,\frac{3}{2},r^2)$ is the wave function
of the bound state in relative coordinates \cite{Busch1998} and
$U(n,m,x)$ is the confluent hypergeometric function and
$\nu=\frac{E}{2 \hbar \omega}-3/4$. We determine the relative
height of the atomic and the molecular peak in the spectra (fig.
\ref{fig3}) and divide the value by the overlap integral. The
resulting value gives the fraction of doubly occupied sites in the
lattice (see fig. \ref{fig5}).

In the tight binding limit we calculate the fraction of doubly
occupied sites of the lattice for a given temperature. For the
analytic calculation we start from the density of states given by
\begin{equation}
\rho_{3D}(E)=\frac{2 \pi E^{1/2}}{\left(\frac{m
\omega^2\lambda^2}{8} \right)^{3/2}}. \label{rho3d}
\end{equation}
and determine the number of doubly occupied sites of a two
component, noninteracting Fermi gas, where both species have the
same Fermi distribution function
$f(E)=\left(e^{(E-\mu)/k_BT}+1\right)^{-1}$ according to
\begin{eqnarray}
N_2&=&\int_{-\infty}^\infty \rho(E) f^2(E) dE. \label{eq1}
\end{eqnarray}
Here the chemical potential $\mu$ is determined from the
normalization to the total particle number
$N=\int_{-\infty}^\infty \rho(E) f(E) dE$ per spin state. In a
three-dimensional optical lattice, we use an expansion similar to
the Sommerfeld expansion for free electrons \cite{Ashcroft1976} to
obtain the fraction of doubly occupied lattice sites
\cite{Kohl2005c}
\begin{eqnarray}
n_2&=&\frac{N_2}{N}=1-\frac{3}{2}\frac{k_B
T}{E_F}+\frac{\pi^2}{8}\left(\frac{k_BT}{E_F}\right)^3+\mathcal{O}\left(\textstyle{(\frac{k_BT}{E_F})}^4\right).
\label{prediction3D}
\end{eqnarray}
From our measured value of $n_2=0.43$ we conclude that the
temperature of the atoms in the optical lattice is at most
$T/T_F=0.46$. A similar result for our experimental data was
computed numerically in \cite{Katzgraber2005}. This value gives an
upper limit to the temperature since it assumes adiabatic
formation of molecules at all doubly occupied sites and a perfect
50:50 mixture of the spin states. Moreover, we have not verified
experimentally that the sample in the lattice is indeed in thermal
equilibrium. During the ramp across the Feshbach resonance the
density distribution might slightly change as compared to the
initial noninteracting case, which limits the accuracy of the
temperature determination. This principle of thermometry can be
also extended to lower potential depth of the optical lattice. For
this regime, however, a full numerical calculation of the double
occupancy is required.

\section{Prospects for interacting fermions in optical lattices}

The experiments on Feshbach resonances in optical lattices have
already demonstrated the physical richness of interacting Fermi
systems. The physics of the Hubbard model can be accessed and
studied. Attaining superfluid, Mott insulating and
anti-ferromagnetic phases \cite{Hofstetter2002} will establish
optical lattices as a tool for quantum simulations of many-body
states. Moreover ``new physics'' can be expected beyond the single
band Hubbard model, a regime which is accessible with Feshbach
resonances in optical lattices. Experimental investigations could
promote an understanding of this regime since quantitative
theoretical predictions are very difficult to obtain.

\section*{Acknowledgements}
We acknowledge funding by SNF, OLAQUI (EU FP6-511057), Qudedis
(ESF), and QSIT.

\section*{References}

\bibliographystyle{unsrt}
\bibliography{JPhysB4}

\begin{thebibliography}{10}

\bibitem{Jaksch1998}
D.~Jaksch, C.~Bruder, J.I. Cirac, C.W. Gardiner, and P.~Zoller.
\newblock Cold bosonic atoms in optical lattices.
\newblock {\em Physical Review Letters}, 81:3108, 1998.

\bibitem{Hofstetter2002}
W.~Hofstetter, J.I. Cirac, P.~Zoller, E.~Demler, and M.D. Lukin.
\newblock High-temperature superfluidity of fermionic atoms in optical
  lattices.
\newblock {\em Physical Review Letters}, 89:220407, 2002.

\bibitem{Hubbard1963}
J.~Hubbard.
\newblock Electron correlations in narrow energy bands.
\newblock {\em Proceedings of the Royal Society of London, Series A}, 276:238,
  1963.

\bibitem{Inouye1998}
S.~Inouye, M.R. Andrews, J.~Stenger, H.-J. Miesner, D.M. Stamper-Kurn, and
  W.~Ketterle.
\newblock Observation of {Feshbach} resonances in a {Bose}-{Einstein}
  condensate.
\newblock {\em Nature}, 392:151, 1998.

\bibitem{Carr2005}
L.~D. Carr and M.~J. Holland.
\newblock Quantum phase transitions in the {Fermi}-{Bose} {Hubbard} model.
\newblock {\em Physical Review A}, 72:031604, 2005.

\bibitem{Zhou2005}
F.~Zhou.
\newblock Mott states under the influence of fermion-boson conversion: invasion
  of superfluidity.
\newblock {\em cond-mat/0505740}.

\bibitem{Diener2006}
R.~B. Diener and T.-L. Ho.
\newblock Fermions in optical lattices swept across {Feshbach} resonances.
\newblock {\em Physical Review Letters}, 96:010402, 2006.

\bibitem{Katzgraber2005}
H.~G. Katzgraber, A.~Esposito, and M.~Troyer.
\newblock Ramping fermions in optical lattices across a {Feshbach} resonance.
\newblock {\em cond-mat/0510194}, 2005.

\bibitem{Feshbach1958}
H.~Feshbach.
\newblock A unified theory of nuclear reactions.
\newblock {\em Annals of Physics}, 5:337, 1958.

\bibitem{Tiesinga1993}
E.~Tiesinga, B.J. Verhaar, and H.T.C. Stoof.
\newblock Threshold and resonance phenomena in ultracold ground-state
  collisions.
\newblock {\em Phyiscal Review A}, 47:4114, 1993.

\bibitem{Busch1998}
T.~Busch, B.-G. Englert, K.~Rzazewski, and M.~Wilkens.
\newblock Two cold atoms in a harmonic trap.
\newblock {\em Foundation of Physics}, 28:549, 1998.

\bibitem{Blume2002}
D.~Blume and C.~H. Greene.
\newblock Fermi pseudopotential approximation: Two particles under external
  confinement.
\newblock {\em Physical Review A}, 65:043613, 2002.

\bibitem{Bolda2002}
E.~L. Bolda, E.~Tiesinga, and P.~S. Julienne.
\newblock Effective-scattering-length model of ultracold atomic collisions and
  {Feshbach} resonances in tight harmonic traps.
\newblock {\em Physical Review A}, 66:013403, 2002.

\bibitem{Dickerscheid2005}
D.B.M. Dickerscheid, U.~Al Khawaja, D.~van Oosten, and H.T.C. Stoof.
\newblock Feshbach resonances in an optical lattice.
\newblock {\em Physical Review A}, 71:043604, 2005.

\bibitem{Gao1998}
B.~Gao.
\newblock Quantum-defect theory of atomic collisions and molecular vibration
  spectra.
\newblock {\em Physical Review A}, 58:4222, 1998.

\bibitem{Troyer2005}
M.~Troyer and U.-J. Wiese.
\newblock Computational complexity and fundamental limitations to fermionic
  quantum {Monte} {Carlo} simulations.
\newblock {\em Physical Review Letters}, 94:0408370, 2005.

\bibitem{Kohl2005b}
M.~K{\"o}hl, H.~Moritz, T.~St{\"o}ferle, K.~G{\"u}nter, and T.~Esslinger.
\newblock Fermionic atoms in a three dimensional optical lattice: {O}bserving
  {Fermi} surfaces, dynamics, and interactions.
\newblock {\em Physical Review Letters}, 94:080403, 2005.

\bibitem{Regal2003b}
C.A. Regal, C.~Ticknor, J.L. Bohn, and D.S. Jin.
\newblock Creation of ultracold molecules from a {Fermi} gas of atoms.
\newblock {\em Nature}, 424:47, 2003.

\bibitem{Greiner2001b}
M.~Greiner, I.~Bloch, O.~Mandel, T.W. H{\"a}nsch, and T.~Esslinger.
\newblock Exploring phase coherence in a {2D} lattice of {Bose}-{Einstein}
  condensates.
\newblock {\em Physical Review Letters}, 87:160405, 2001.

\bibitem{Stoferle2005b}
T.~St{\"o}ferle, H.~Moritz, K.~G{\"u}nter, M.~K{\"o}hl, and T.~Esslinger.
\newblock Molecules of fermionic atoms in optical lattices.
\newblock {\em Physical Review Letters}, 96:030401, 2006.

\bibitem{Regal2004a}
C.A. Regal, M.~Greiner, and D.S. Jin.
\newblock Observation of resonance condensation of fermionic atom pairs.
\newblock {\em Physical Review Letters}, 92:040403, 2004.

\bibitem{Chin2004}
C.~Chin, M.~Bartenstein, A.~Altmeyer, S.~Riedl, S.~Jochim, J.~Hecker Denschlag,
  and R.~Grimm.
\newblock Observation of the pairing gap in a strongly interacting {Fermi} gas.
\newblock {\em Science}, 305:1128, 2004.

\bibitem{Moritz2005}
H.~Moritz, T.~St{\"o}ferle, K.~G{\"u}nter, M.~K{\"o}hl, and T.~Esslinger.
\newblock Confinement induced molecules in a {1D} {Fermi} gas.
\newblock {\em Physical Review Letters}, 94:210401, 2005.

\bibitem{Wigner1933}
E.~Wigner.
\newblock {{\"U}ber die Streuung von Neutronen an Protonen}.
\newblock {\em Zeitschrift f{\"u}r Physik}, 83:253, 1933.

\bibitem{Harber2002}
D.M. Harber, H.J. Lewandowski, J.M. McGuirk, and E.A. Cornell.
\newblock Effect of cold collisions on spin coherence and resonance shifts in a
  magnetically trapped ultracold gas.
\newblock {\em Physical Review A}, 66:053616, 2002.

\bibitem{Gupta2003}
S.~Gupta, Z.~Hadzibabic, M.W. Zwierlein, C.A. Stan, K.~Dieckmann, C.H. Schunck,
  E.G.M. van Kempen, B.J. Verhaar, and W.~Ketterle.
\newblock Radio-frequency spectroscopy of ultracold fermions.
\newblock {\em Science}, 300:1723, 2003.

\bibitem{Regal2003a}
C.A. Regal and D.S. Jin.
\newblock Measurement of positive and negative scattering lengths in a {Fermi}
  gas of atoms.
\newblock {\em Physical Review Letters}, 90:230404, 2003.

\bibitem{Regal2004b}
C.A. Regal, M.~Greiner, and D.S. Jin.
\newblock Lifetime of molecule-atom mixtures near a {Feshbach} resonance in
  {$^{40}$K}.
\newblock {\em Physical Review Letters}, 92:083201, 2004.

\bibitem{Rabl2003}
P.~Rabl, A.J. Daley, P.O. Fedichev, J.I. Cirac, and P.~Zoller.
\newblock Defect-suppressed atomic crystals in an optical lattice.
\newblock {\em Physical Review Letters}, 91:110403, 2003.

\bibitem{Blakie2005}
P.~B. Blakie and A.~Bezett.
\newblock Adiabatic cooling of fermions in an optical lattice.
\newblock {\em Physical Review A}, 71:033616, 2005.

\bibitem{Werner2005}
F.~Werner, O.~Parcollet, A.~Georges, and S.~R. Hassan.
\newblock Interaction-induced adiabatic cooling and antiferromagnetism of cold
  fermions in optical lattices.
\newblock {\em Physical Review Letters}, 95:056401, 2005.

\bibitem{Ashcroft1976}
N.W. Ashcroft and N.D. Mermin.
\newblock {\em Solid State Physics}.
\newblock Harcourt College Publishers, 1976.

\bibitem{Kohl2005c}
M.~K{\"o}hl.
\newblock Thermometry of fermionic atoms in optical lattices.
\newblock {\em Physical Review A}, 73:031601, 2006.

\end{thebibliography}

\end{document}